\documentclass[nofootinbib,prl,twocolumn,aps,preprintnumbers]{revtex4-2}
\usepackage{epsfig,graphics}
\usepackage{amsfonts,amssymb,amsmath}            
\usepackage{ifthen}
\usepackage{amsthm}
\usepackage{hyperref}
\usepackage{url}
\usepackage{todonotes}
\usepackage{color, colortbl,hhline}
\usepackage{tikz}
\usetikzlibrary{matrix}
\usepackage{diagbox}

\newcommand{\comment}[1]{}
\newcommand{\ket}[1]{| #1 \rangle}
\newcommand{\bra}[1]{\langle #1 |}

\newcommand{\bs}[1]{\boldsymbol{#1}}

\theoremstyle{plain}
\newtheorem{theorem}{Theorem}

\theoremstyle{definition}

\definecolor{amber}{rgb}{1.0, 0.75, 0.0}
\definecolor{aureolin}{rgb}{0.99, 0.93, 0.0}
\definecolor{steel}{HTML}{91CDE2}
\definecolor{bblue}{HTML}{508AA8}
\definecolor{beyes}{HTML}{EDF6FC}
\newcommand{\n}[1]{|#1|}

\newcommand{\mc}[1]{\mathcal{#1}}
\newcommand{\mt}[1]{\mathrm{#1}}

\begin{document}

\title{Entanglement cost for steering assemblages}
\date{\today}

\author{Thomas \surname{Cope}}
\email{thomas.cope@itp.uni-hannover.de}
\address{Institut f{\"u}r Theoretische Physik, Leibniz Universit{\"a}t Hannover, Appelstr. 2, 30167 Hannover, Germany}

\begin{abstract}
In this paper we answer the question of how one can \emph{quantify} the amount of entanglement necessary to create a given steering assemblage, when the original state and measurements are unknown. To do this, we extend the concepts of \emph{entanglement cost} and \emph{entanglement of formation} to steering assemblages, and provide easy to calculate upper and lower bounds. We prove that the entanglement of formation for assemblages is not generally continuous and not a flat roof extension (in contrast to the state-based quantity); and use numerical analysis to illustrate these properties. Finally, we discuss the consequences of these results for assemblage-to-assemblage conversion and suggest potential applications.
\end{abstract}
\maketitle



Quantum steering is a quickly growing field in quantum information research, although its conception can be traced far back into quantum mechanics' history. Originally explored by Schr{\"o}dinger \cite{S1935,S1936} to understand the consequences of the EPR paradox \cite{EPR1935}, it can be summarised as follows: measurement on one part of an entangled state can cause the other part of the state to be ``steered" into a preferred decomposition. Although this cannot be used to transmit information, this is nevertheless a surprising result, described by Schr{\"o}dinger himself as both ``disconcerting" and ``discomforting" \cite{S1936}.

Interest in quantum steering was revived in 2007 in \cite{WJD2007}, where the modern definition was formalised by defining it as the negation of a ``local hidden state" model; similarly to how entanglement is the negation of separable states, and Bell nonlocality the negation of a ``local hidden variable" model. Steering is often considered ``halfway" between entanglement, which doesn't consider measurements at all, and nonlocality, which is concerned \emph{only} with the outcomes of measurements. Therefore, steering inherits properties from both of its near neighbours. Steering is also operationally relevant to \emph{one-sided} (or semi) Device Independent Quantum Key Distribution (DIQKD) \cite{BCWSW2012,WBLZL2013,KWW2020}, for which steering is the natural resource theory \cite{GA2015}. In this scenario, one device is treated as a block box, while the other performs trusted quantum processes. As characterising a device is both expensive in terms of knowledge and resources, such one-sided protocols are useful between users of asymmetric capital e.g. a bank and its many customers.\\ 
It is known that in order to exhibit the phenomenon of steering, one must measure an entangled state using incompatible measurements. Thus, observing the phenomenon of steering implicitly verifies entanglement of the underlying state. In this paper the goal is to \emph{quantify} that verification, by asking how much entanglement is required to prepare a system exhibiting steering.
\section{Assemblages and Entanglement}
Mathematically, steering is described via assemblages: a set of subnormalised substates $\{\sigma_{n|r}\}_{n,r}$ such that
\begin{equation}
\sigma_{n|r}:=\mathrm{Tr}_A\left(M^A_{n|r}\otimes \mathbb{I}^B\rho^{AB}\right),\label{AssemCreate}
\end{equation}
where $\rho^{AB}$ is a bipartite quantum state, and $M_{n|r}$ is an element of the Postive Operator Valued Measure (POVM) $\{M_{n|r}\}_{n}$ i.e. $\sum_n M_{n|r}= \mathbb{I}^A$. Normally, we fix the number of measurements Alice can perform, and the number of outputs she can have, so that $r\in \mc{R}=\{0,\ldots |\mc{R}|-1\}$ and $n\in\mc{N}=\{0,\ldots |\mc{N}|-1\}$. From this definition come the natural constraints 
\begin{equation}
\sigma_{n|r}\geq 0\; \forall n,r \;\;\;\; \sum_{n}\sigma_{n|r}=\mathrm{Tr}_{A}(\rho^{AB})\,=:\rho_{\bs{\sigma}}^{B},\label{SteerCon}
\end{equation}
known as \emph{positivity} and \emph{nosignalling} respectively. From now on we will write an assemblage succinctly as $\boldsymbol{\sigma}:= (\sigma_{0|0},\ldots ,\sigma_{(|\mc{N}|-1)|0},\ldots ,\sigma_{0|(|\mc{R}|-1)},\ldots ,\sigma_{(|\mc{N}|-1)|(|\mc{R}|-1)})$. We will refer to the marginal state of this assemblage as $\rho_{\boldsymbol{\sigma}}$.
In this paper, we are limiting our discussion to finite (though arbitrary dimension) Hilbert spaces. This will allow us to use tricks such as Schmidt decomposition and purification. \\
In the case of bipartite systems, the mapping from all quantum states and measurements to assemblages is surjective, but not injective; the entangled Werner state
\begin{equation}
W_{1/\sqrt{2}}=\frac{1}{\sqrt{2}}\ket{\Phi^+}\bra{\Phi^+} + \left(1-\frac{1}{\sqrt{2}}\right)\frac{\mathbb{I}}{4}
\end{equation}
(where $\Phi^{+}=\left(\ket{00}+\ket{11}\right)\sqrt{2}$), along with measurements
\begin{align*}
M_{0|0}&=\ket{0}\bra{0},&  M_{0|1}&=\ket{+}\bra{+} ,\\
M_{1|0}&=\ket{1}\bra{1}, & M_{1|1}&=\ket{-}\bra{-},
\end{align*}
creates the assemblage
\begin{align*}
\sigma_{0|0}&=\frac{1}{2}\left(\mathbb{I}+\frac{1}{\sqrt{2}}\sigma_z\right)
 ,& \sigma_{0|1}&=\frac{1}{2}\left(\mathbb{I}+\frac{1}{\sqrt{2}}\sigma_x\right),\\
\sigma_{1|0}&=\frac{1}{2}\left(\mathbb{I}-\frac{1}{\sqrt{2}}\sigma_z\right)
 ,& \sigma_{1|1}&=\frac{1}{2}\left(\mathbb{I}-\frac{1}{\sqrt{2}}\sigma_x\right).\\
\end{align*}
However, so too does the separable state 
\begin{equation}
\rho^{\mathrm{Sep}}=1/4\sum_{i,j=0}^{1} \ket{ij}^{A}\bra{ij} \otimes \sigma_{ij}^{B},
\end{equation}
where $ \sigma_{ij}=\frac{1}{2}\left(\mathbb{I}+ \frac{(-1)^j}{\sqrt{2}}\sigma_x + \frac{(-1)^i}{\sqrt{2}}\sigma_z\right)$, using measurements
\begin{align*}
M^{A}_{n|0}&= \ket{n}\bra{n}\otimes \mathbb{I},& M^{A}_{n|1}&= \mathbb{I}\otimes \ket{n}\bra{n}.
\end{align*}
This example in particular highlights the fact that the same assemblage can arise from states with a different quantity of entanglement. 
 This leads to a natural question: \emph{how much entanglement is necessary to create a given assemblage} $\bs{\sigma}$? To answer this we define the \emph{entanglement cost for assemblages},

\begin{align}
	E_{\mt{CA}}(\bs{\sigma}):= \inf \bigg\{ &E \;\bigg\vert\; \forall\, \epsilon > 0, \delta > 0, \exists\, p,q,\nonumber \\
	&\forall\, \mathbf{n}\in\mc{N}^{\times q}, \mathbf{r}\in\mc{R}^{\times q},\,\exists\, M_{\bs{n}|\bs{r}}\geq 0,\,\Lambda\nonumber\\
	&\sum_{\bs{a}} M_{\bs{n}|\bs{r}}=\mathbb{I}^{A}, \; |E-\frac{p}{q}|\leq \delta\; \text{and} \nonumber \\
	\sum_{\bs{n},\bs{r}} \big\|\sigma_{\bs{n}|\bs{r}}-\mathrm{Tr}_{A}&\left[\left(M_{\bs{n}|\bs{r}}\otimes \mathbb{I}^{B}\right)\Lambda(\ket{\Phi^+}\bra{\Phi^+}^{\otimes p})\right]\big\| \leq \epsilon\bigg\}.
\end{align}
In this formula $\Lambda$ is a Local Operations and Classical Communications (LOCC) distillation procedure, used to create a shared state between Alice and Bob. Alice then measures her system (which may be of arbitrary finite size) with a measurement whose operators are given by $M_{\bs{n}|\bs{r}}$, with the goal of creating an approximation to $q$ copies of the steering assemblage. We represent $q$ copies of the steering assemblage by a larger assemblage, in which Alice may choose as input any $q$ tuple $\mathbf{r}\in \mc{R}^{\times q}$, representing the inputs of each individual component. The possible substates which can occur as a result of this choice consist of all combinations of possible substates output by the individual smaller assemblages, in such a way that the outputs are independent of each other. Thus a possible substate is represented by
\begin{equation}
\sigma_{\bs{n}|\bs{r}} = \sigma_{n_1|r_1} \otimes \sigma_{n_2|r_2} \otimes \ldots \otimes \sigma_{n_q|r_q}.
\end{equation}
A more detailed treatment of this is given in the supplementary material. The above definition is designed in mind that the entanglement cost for creating multiple copies of an assemblage may be lower than that of a single assemblage. Furthermore, we allow the freedom for Alice and Bob to create ``good approximations" of the assemblage, such that in the asymptotic limit these approximations become exact. This is directly analogous to the entanglement cost for quantum states,
\begin{align}
E_{\mt{C}}(\rho):= \inf \bigg\{ &E \mid \forall \epsilon > 0, \delta > 0, \exists m, n, \Lambda, |E-\frac{p}{q}| \leq \delta \nonumber \\
& \text{ and } \|\rho^{\otimes q}-\Lambda(\ket{\Phi^+}\bra{\Phi^+}^{\otimes p})\| \leq \epsilon\bigg\},
\end{align}
where $\Lambda$ is again a LOCC distillation procedure.\\
One key difference between the two definitions is that, when creating assemblages, it may be that the measurements are only asymptotically correct, instead of (or as well as) the state. We also address in the supplemental material, why no post-processing term $\Lambda'$ is included after the measurement.\\
 Calculation of $E_{\mt{C}}(\rho)$ has been shown to be NP-complete \cite{H2014}, a property we conjecture carries over to $E_{\mt{CA}}$ due to the definition. However, we do know its value for some special cases: when the assemblage has a Local Hidden State (LHS) model of the form \footnote{Normally this condition is expressed as an integral over the hidden variable $\lambda$, however, for fixed $n,r$ the form written here is entirely equivalent.}
\begin{equation}
\sigma^{\mathrm{LHS}}_{n|r}=\sum_{\lambda=1}^{d} D(n|\lambda,r)\sigma_{\lambda},
\end{equation}
where $D(n|\lambda,r)$ describes a possible deterministic assignment of outputs to all inputs, with $d$ the total number of these, and $\sigma_{\lambda} \geq 0$. 
Then the entanglement cost is
\begin{equation}
E_{\mt{CA}}(\bs{\sigma}^{\mathrm{LHS}})=0.
\end{equation}

This follows from the fact that every LHS assemblage, also known as \emph{unsteerable} assemblages, can be created from a separable state  \cite{KSCAA2015,MGHUG2016}, which can be constructed explicitly via a semidefinite program \cite{CS2017}. It is also worth noting that all assemblages created from separable states are unsteerable. 
We can always upper bound $E_{\mt{CA}}(\bs{\sigma}) \leq S(\rho_{\bs{\sigma}})$ i.e. by the von Neumann entropy of the marginal state  $S(\rho_{\bs{\sigma}}):=-\mathrm{Tr}(\rho_{\bs{\sigma}}\log \rho_{\bs{\sigma}})$. This is because we have an explicit realisation of the assemblage which is constructed from the marginal $\rho_{\bs{\sigma}} = \sum_{i}\lambda^i \ket{v^i}\bra{v^i}$ via

\begin{align*}
\rho^{AB}&=\sum_{i}\sqrt{\lambda^i} \ket{v^i}^A\ket{v^i}^B,\\
M_{n|r}&= \rho_{\bs{\sigma}}^{-1/2}\sigma_{n|r}^{T}\rho_{\bs{\sigma}}^{-1/2},
\end{align*}

where transposition is taken with respect to the basis $\{\ket{v^i}\}^i$, and the pseudo-inverse is used if necessary. Putting these into Eq. (\ref{AssemCreate}) will reproduce our desired assemblage.\\
Since for pure states $E_{\mt{C}}(\ket{\Phi}^{AB}\bra{\Phi})= S(\mathrm{Tr}_A(\ket{\Phi}^{AB}\bra{\Phi})$, we have  that $E_{\mt{C}}(\rho^{AB})=S(\rho_{\bs{\sigma}})$. This provides an upper bound to $E_{\mt{CA}}(\boldsymbol{\sigma})$. However, in general it provides quite a bad lower bound; for our earlier example, which we know has a value of $E_{\mt{CA}}=0$, it would give the upper bound $\leq 1$. We want to do better than this.\\
Instead of considering the entanglement cost directly, we propose another measure: the \emph{entanglement of formation for assemblages}
\begin{align}
E_{\mt{FA}}(\bs{\sigma}):=\inf \bigg\{ &\sum_{i} p^i S(\rho_{\bs{\sigma^i}})\;|\; \sum_{i} p^i \bs{\sigma}^i =\bs{\sigma},\; p^i\geq 0, \nonumber\\
&\sum_i p^i=1,\;\bs{\sigma}^i \text{ extremal in } \mathcal{G}_{\mathcal{N}|\mathcal{R}} \bigg\}
\end{align}
where $\mathcal{G}_{\mathcal{N}|\mathcal{R}}$ is the set of all assemblages with inputs in $\mathcal{R}$  and outcomes in $\mathcal{N}$. One can verify from the relations in Eq. (\ref{SteerCon}) this set is convex. The assemblages $\bs{\sigma}^i$ are \emph{extremal} points; that is, they can only be written as trivial convex combinations of themselves \cite{CO2021}. \\
This is a direct analogy with the \emph{entanglement of formation} for quantum states \cite{BDSW1996}
\begin{align}
	E_{\mt{F}}(\rho):=\inf \bigg\{ &\sum_{i} p^i S(\mathrm{Tr}_{A}(\ket{\phi^i}\bra{\phi^i}))\;|\; \sum_{i} p^i \ket{\phi^i}\bra{\phi^i} =\rho,\;\nonumber\\
	 &p^i\geq 0, \;
	\sum_i p^i=1\bigg\},
\end{align}
which is an upper bound $E_{\mt{F}}(\rho)\geq E_{\mt{C}}(\rho)$. Moreover, the stronger relation holds that $\lim_{ n\rightarrow \infty} E_{\mt{F}}(\rho^{\otimes n})/n=E_{\mt{C}}(\rho)$ \cite{HHT2001}.\\
Both the original entanglement of formation and the entanglement of formation for assemblages are examples of \emph{convex roof extensions} \cite{U2010}. These consist of a convex set $\mathcal{X}$ with a function $g(x)$ defined on the boundary ($\partial \mathcal{X}$).  
This is then extended to a function on the entire set ($G(x)$) by 
{\small
\begin{align}
&G(x):=\nonumber\\
&\inf \left\{\sum_i g(x^i)\mid \sum_i p^i x^i = x,\; \sum_i p^i = 1,\; p^i\geq 0,\; x^i\in \partial \mc{X}\right\}.
\end{align}
}
The convex roof extension is equivalently defined as the largest convex function $F$ on $\mc{X}$ such that $F(x)=g(x)$ when $x\in \partial \mathcal{X}$. Another important result is that for the entanglement of formation for quantum states (along with many other convex roof extensions), the optimal decomposition satisfies $G(x)=g(x^i),\, \forall i$.\\
Why do we choose to define the entanglement of formation for assemblages this way? To see this, first we show that  $S(\rho_{\boldsymbol{\sigma}})$ corresponds to the entanglement for extremal assemblages.\qed
\begin{theorem}
If $\boldsymbol{\sigma}$ is extremal, then any state $\rho^{AB}\in \mc{S}_{\bs{\sigma}}$ must have $E_{\mt{F}}(\rho^{AB})=S(\rho_{\boldsymbol{\sigma}})$.\\
\end{theorem}
If $\rho^{AB}$ is pure, i.e. $=\ket{\phi}\bra{\phi}$, then this is true straightaway, as $E_{\mt{F}}(\ket{\phi}\bra{\phi})=E_{\mt{C}}(\ket{\phi}\bra{\phi})=S(\mathrm{Tr}_{A}\ket{\phi}\bra{\phi})=S(\rho_{\boldsymbol{\sigma}})$. If the state is mixed, i.e. $\rho^{AB}=\sum_{i} p^i \ket{\phi^i}\bra{\phi^i}$, one can decompose $\bs{\sigma}$ as
\begin{equation}
\bs{\sigma}=\sum_{i} p^i\bs{\sigma}^{i}\;\;\;\sigma_{n|r}^{i}:=\mathrm{Tr}_A\left(M^A_{n|r}\otimes \mathbb{I}^B\ket{\phi^i}\bra{\phi^i}\right).
\end{equation}
However, our assemblage is extremal, and thus $\boldsymbol{\sigma}^{i}=\boldsymbol{\sigma},\; \forall i$. In particular, this forces all $\ket{\phi^i}$ to have the marginal $\rho_{\boldsymbol{\sigma}}$. Thus the entanglement of $\ket{\phi^i}$ is $S(\rho_{\boldsymbol{\sigma}})$, and since this holds for all pure-state decompositions of $\rho^{AB}$, $E_{\mt{F}}(\rho^{AB}) = S(\rho_{\boldsymbol{\sigma}})$.

\begin{theorem}\label{Thm2}
$E_{\mt{FA}}(\bs{\sigma})=\min \left\{ E_{\mt{F}}(\rho^{AB}) \mid \rho^{AB}\in\mathcal{S}_{\bs{\sigma}} \right\}$, where $\mathcal{S}_{\bs{\sigma}}$ is the set of all states such that $\exists$ POVMs with $\sigma_{n|r}=\mathrm{Tr}_{A}\left[\left(M_{n|r} \otimes \mathbb{I}^{B}\right)\rho^{AB}\right]$.
\end{theorem}
First we prove that for all $\rho^{AB}\in \mathcal{S}_{\bs{\sigma}} $ there exists a decomposition $\bs{\sigma}=\sum_i p^i\bs{\sigma}^{i}$ such that $\sum_i p^i S(\rho_{\bs{\sigma}^{i}})=E_{\mt{F}}(\rho^{AB})$. This is done simply by decomposing
\begin{equation}
\bs{\sigma}=\sum_i p^i \bs{\sigma}^i,\;\sigma_{n|r}^{i}:=\mathrm{Tr}_A\left(M^A_{n|r}\otimes \mathbb{I}^B\ket{\phi^i}\bra{\phi^i}\right),
\end{equation}
where $\ket{\phi^i}$ are in the optimal decomposition of $\rho^{AB}$. This gives us that
\begin{equation}
E_{\mt{FA}}(\bs{\sigma}) \leq \sum p^i S(\mathrm{Tr}_A(\ket{\phi^i}\bra{\phi^i}) = E_{\mt{F}}(\rho_{AB}).
\end{equation}
This holds even if $\bs{\sigma}^i$ are not extremal, since the von Neumann entropy is concave. This implies that $E_{\mt{FA}}(\bs{\sigma})\leq \min E_{\mt{F}}(\rho_{AB})$.

Conversely, suppose there exists a decomposition $\boldsymbol{\sigma}=\sum_{i} p^i \boldsymbol{\sigma}^{i}$ such that $\sum_{i} p^i S(\boldsymbol{\sigma}^{i}) = E$. \\
Let us define the corresponding  pure state and measurements creating $\boldsymbol{\sigma}^{i}$ as $\ket{\phi^i}\bra{\phi^i}$, $M_{a|x}^{i}$. Then we can construct $\boldsymbol{\sigma}$ via the state and measurements
\begin{align*}
\rho^{A'AB}&:=\sum_{i} \ket{i}^{A'}\bra{i}\otimes \ket{\phi^i}\bra{\phi^i},\\
M^{A'A}_{n|r}&:=\sum_{i} \ket{i}^{A'}\bra{i}\otimes M_{n|r}^{i}.
\end{align*}
$\rho^{A'AB}$ is decomposable into pure states $\ket{i}\bra{i}\otimes \ket{\phi^i}^{AB}\bra{\phi^i}$, whose entanglement (with respect to the bipartition $AA'|B$) is $S(\mathrm{Tr}_A\ket{\phi^i}^{AB}\bra{\phi^i})=S(\bs{\sigma}^i)$. Thus $E_{\mt{F}}(\rho^{A'AB})\leq E$, implying $\min E_{\mt{F}}(\rho^{AB}) \leq E_{\mt{FA}}(\bs{\sigma})$ (note there is no restriction of the dimension of $A$).\qed

Theorem \ref{Thm2} gives us a direct connection between $E_{\mt{FA}}(\bs{\sigma})$ and the entanglement of formation for quantum states. Since for all states $E_{\mt{C}}(\rho)\leq E_{\mt{F}}(\rho)$, we see that it also gives an upper bound $E_{\mt{CA}}(\bs{\sigma})\leq E_{\mt{FA}}(\bs{\sigma})$. An important question is whether there exists an assemblage which is optimally created from a state satisfies $E_{\mt{F}}=E_{\mt{C}}$, but where $E_{\mt{CA}}(\bs{\sigma}) < E_{\mt{FA}}(\bs{\sigma})$. 
	
For this measure we may state that $E_{\mt{FA}}(\bs{\sigma})=0$ iff $\bs{\sigma}$ is unsteerable. For an unsteerable assemblage $\bs{\sigma}$, we may find the corresponding separable state by semidefinite programming - decomposing this into pure states will then the desired extremals $\bs{\sigma}^{i}$. Conversely, any steering assemblage created from a separable state is unsteerable \cite{CS2017}. Furthermore, one can always upper bound $E_{\mt{FA}}(\bs{\sigma})$ by use of the \emph{steering weight}, a steering monotone calculable by semidefinite programming. The steering weight is defined as \cite{SNC2014}
	\begin{equation}
		SW(\bs{\sigma}):= \min\{\;p\; | \bs{\sigma}=p\bs{\gamma} + (1-p)\bs{\sigma}^{\mathrm{LHS}}\},
	\end{equation}
	i.e. a convex combination of a steerable and unsteerable assemblage. This gives the upper bound
\begin{equation}
	E_{\mt{FA}}(\bs{\sigma})\leq p S(\rho_{\bs{\gamma}}):= E_{\mathrm{SW}}(\bs{\sigma}).
\end{equation}
We can also provide a lower bound for $E_{\mt{CA}}$ (and thus $E_{\mt{FA}}$) in the form of the \emph{distillable secret key}, $K_{\mt{D}}$ \cite{DHR2002,HHHO2005}. To see this, first note that any key distillation protocol from $\bs{\sigma}$ is also a key distillation protocol from $\ket{\Phi^+}$, since we may asymptotically create $1/E_{\mt{CA}}(\bs{\sigma})$ copies of $\bs{\sigma}$ from one copy of $\ket{\Phi}^+$. Therefore if $K_{\mt{D}}/(E_{\mt{CA}}) > 1$ would represent a key distillation rate of $\ket{\Phi^+}$ larger than 1. However, we know that $K_D(\ket{\Phi^+})=1$.\\
In order to lower bound this rate, we use a local entropic uncertainty relation \cite{BCCRR2010,TR2011}, where Alice and Bob are able to make measurements on some state, and Bob's measurements are characterised. This results in a state-independent bound
\begin{equation}
	K_{\mt{D}} \geq -\log_2 c  - H(N'|N) - H(\tilde{N}'|\tilde{N}).\label{LBound}
\end{equation}
In this formula $N$ and $\tilde{N}$ are uncharacterised measurements performed by Alice, whilst $N', \tilde{N}'$ are approximations of $N$ and $\tilde{N}$ respectively, based on known POVMs $\{M^{B,N'}_{n'}\}$, $\{M^{B,\tilde{N}'}_{\tilde{n}'}\}$ performed by Bob.
The value $c$ serves as the \emph{incompatibility} of the two POVMs, and is defined as
\begin{equation}
c = \max_{n',\tilde{n}'} \left\|\sqrt{M^{B,N'}_{n'}}\sqrt{M^{B,\tilde{N}^{\prime}}_{\tilde{n}'}}\right\|^2_{\infty}.\label{incompat}
\end{equation}

One can see that the security of this protocol comes from Bob being able to perform exactly his measurements on a generally unknown entangled state, which allows Bob to correlate two incompatible measurements to Alice. As has been noted in e.g. \cite{BCWSW2012,ML2012}, this is well suited to steering scenarios as Bob can characterise his system, whilst Alice cannot. To make the association with steering clearer, we are associating with the distribution $N$ with $P_{N|r}$ i.e. the distribution over outcomes given Alice's specific choice of $r$, and likewise $\tilde{N}$ represents $P_{N|\tilde{r}}$. To respect this, we shall express Bob's POVM choices as $\{M^{B}_{n'|r}\}$, $\{M^{B}_{\tilde{n}'|\tilde{r}}\}$.\\
Any choice of $\{M^{B}_{n'|r}\}$, $\{M^{B}_{\tilde{n}'|\tilde{r}}\}$ provides a lower bound on $K_{\mt{D}}(\bs{\sigma})$ 
 although it might not be a very tight one. In order to gain an intuition about what a sensible choice might be, let us consider the role of $M^{B}_{n'|r}$. Alice performs measurement $r$ and obtains outcome $n$ with probability $p_{n|r}=\mathrm{tr}(\sigma_{n|r})$. As a result Bob holds the state   $\hat{\sigma}_{n|r}=\sigma_{n|r}/\mathrm{tr}(\sigma_{n|r})$. The aim of $\{M^{B}_{n'|r}\}$ is to determine which state Bob holds and give $n'= n$, given that he receives state $\hat{\sigma}_{n|r}$ with probability $p_{n|r}$. This is a more generally studied task in quantum information known as \emph{quantum state discrimination} \cite{BK2015}. For the case when $|\mc{N}|=2$, the optimal measurement is known, while for $|\mc{N}|>2$ the ``pretty good measurement" \cite{HW1994,B1996,SKMH1998} is a reasonable choice. We also use the Fano inequality,
\begin{equation}
H(N'|N)\leq H_{2}(P(\neq_{r}))+P(\neq_{r})\log_2(|\mc{N}|-1),
\end{equation}
where $P(\neq_{r})=P(N'\neq N)$. This simplifies our lower bound to
\begin{align}
 \max_{r,\tilde{r}} -\log_2 c  &- \left(H_2(P(\neq_{r})) + P(\neq_{r})\log_2(|\mc{N}|-1)\right)\nonumber\\ 
 &- \left(H_2(P(\neq_{\tilde{r}})) + P(\neq_{\tilde{r}})\log_2(|\mc{N}|-1)\right).\label{LBound2}
\end{align}
For $|\mc{N}|=2$, the Fano inequality is tight, and we use
\begin{align*}
M^{B}_{0|r}&=\mathbb{P}_+(\sigma_r), & M^{B}_{1|r}&=\mathbb{P}_-(\sigma_r), &
P(\neq_{r})=\frac{1}{2}-\frac{1}{2}\|\sigma_{r}\|_1,
\end{align*}
where $\mathbb{P}_{+(-)}(\sigma_r)$ is the projector onto the positive (negative) eigenspaces of $\sigma_r:=\sigma_{0|r}-\sigma_{1|r}$.
When $|\mc{N}|>2$, we use
\begin{align}
M^{B}_{n'|r}&=\rho_{\bs{\sigma}}^{-1/2}\sigma_{n'|r}\rho_{\bs{\sigma}}^{-1/2},\nonumber\\
 P(\neq_{r})&=1-\sum_{n'}\mathrm{Tr}\left[\left(\sigma_{n'|r}\rho_{\bs{\sigma}}^{-1/2}\right)^2\right].
\end{align}

\begin{figure} 
	\begin{center}
\includegraphics[scale=0.35]{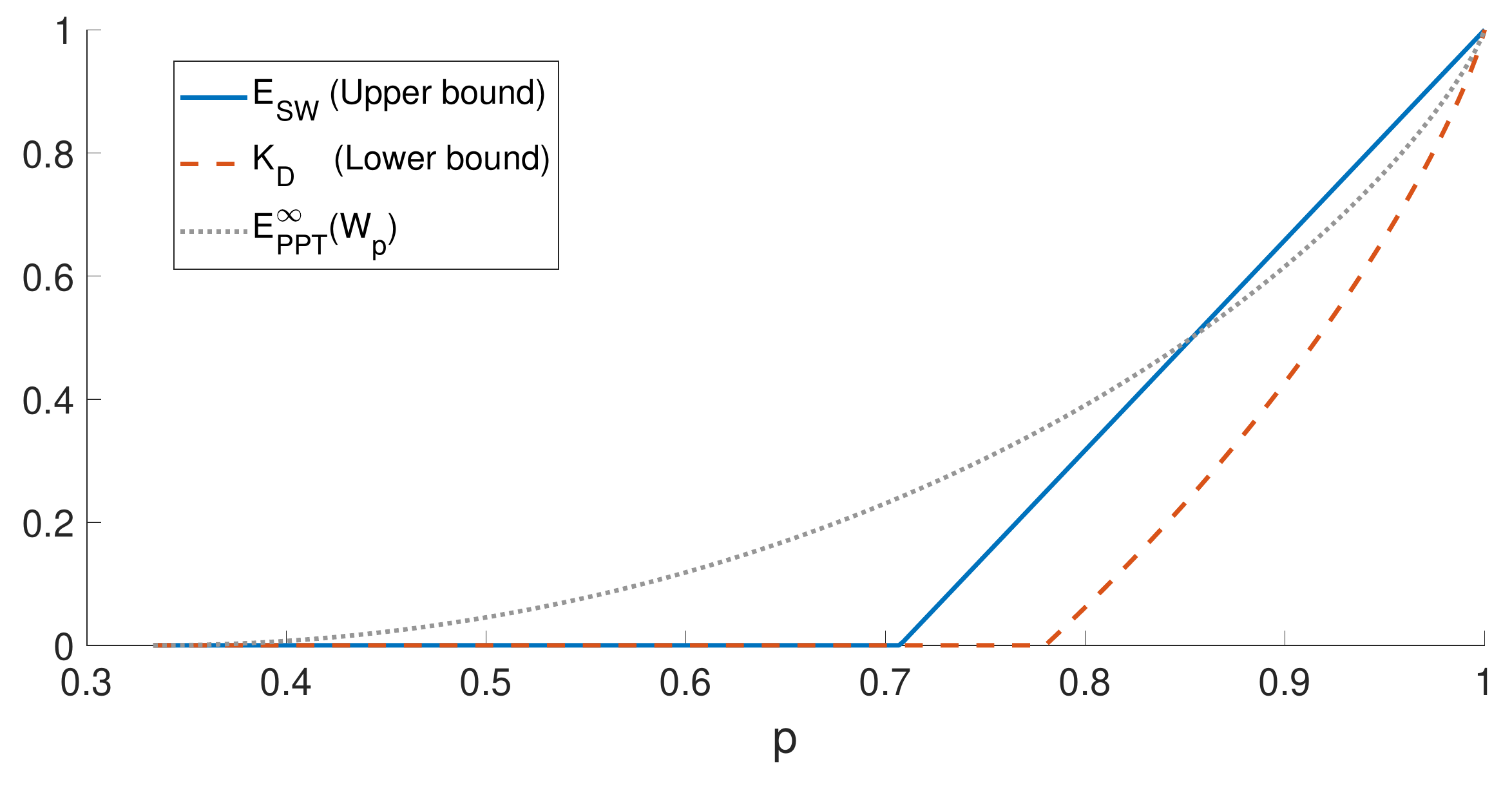}
\end{center}
\caption{Bounds for the entanglement cost of assemblages created by measuring an entangled Werner state $W_{p}=p\ket{\Phi^+}\bra{\Phi^+}+(1-p)\mathbb{I}/4$, with $1/3 < p \leq 1$, in the Pauli-X and Pauli-Z bases. Also plotted is the \emph{Regularized Relative Entropy of Entanglement} with respect to PPT states ($E_{\mathrm{PPT}}^{\infty}$) for Werner states \cite{AEJPVD2001}. As $E_{\mathrm{PPT}}^{\infty}$ is a lower bound to the entanglement cost, we can see that Werner states \emph{cannot} be optimal for $p<0.85406$.}
\end{figure}

\section{Properties of $\mathbf{E_{\mt{FA}}}$ and Numerical Results}
Currently we can only calculate bounds on $E_{\mt{FA}}(\bs{\sigma})$; either by the steering weight or by approximation of extremal assemblages by a finite set, as can be done with computations of the entanglement of formation for states \cite{AVD2001}. Nevertheless, we can say something about its properties:
\begin{itemize}
\item it is not a continuous function;
\item the optimal decomposition, in general, does not consist of $\bs{\sigma}^i$ with equal $S(\rho_{\bs{\sigma}^i})$.
\end{itemize}
The second is in direct contrast to the case for quantum states, a property which was used to prove the remarkable formula for the 2-qubit entanglement of formation \cite{W1998}.\\
To prove these two statements, we turn to the specific case of $|\mc{N}|=|\mc{R}|=2$ and $\mathrm{dim}(\mathcal{H}_B)=2$. For this set, extremal assemblages take the form
\begin{align}
	\sigma_{n|r} = p_{n|r}\ket{\phi_{n|r}}\bra{\phi_{n|r}}\; \forall n,\,r\;\;\nonumber\\ \{\ket{\phi_{0|0}},\ket{\phi_{1|0}}\}\neq\{\ket{\phi_{0|1}},\ket{\phi_{1|1}}\}.
\end{align}
This is best seen by considering the ``slice" of the Bloch sphere considering the points  $\{\ket{\phi_{n|r}}\}$; we see there is only one unique set of weights such that the nosignalling condition holds.\\
\begin{figure} 
	\begin{center}
\includegraphics[scale=0.5]{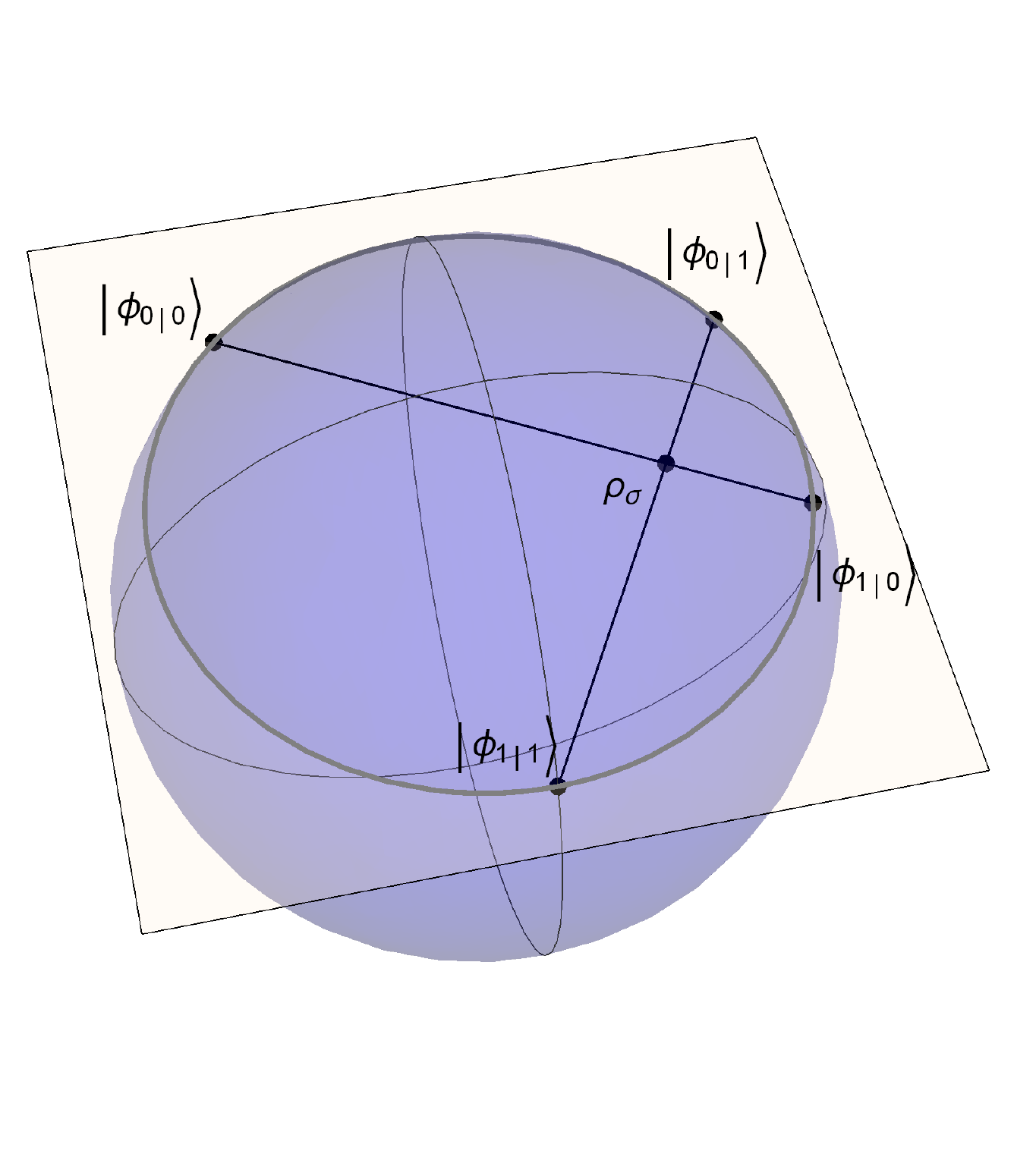}
			
	\caption{An example of an extremal assemblage: the weights of the subnormalised states are given by the relative length to the crossing point, which represents the marginal $\rho_{\bs{\sigma}}$. Only one crossing of the two lines is possible, and the individual points cannot be decomposed; these two properties makes the assemblage extremal.}
	\end{center}
\end{figure}
There are also  extremal assemblages of the form
\begin{equation}
	\sigma_{n|r}=\delta_{nn_r}\ket{\phi}\bra{\phi}.
\end{equation}
which correspond to embeddings of $|\mc{N}|=1,\, |\mc{R}|=2$ into the space.\\
The noncontinuity of $E_{\mt{FA}}$ can seen by considering the assemblage 
\begin{align*}
	\sigma^{\epsilon}_{0|0}&=\frac{1}{2}\ket{0}\bra{0},
	& \sigma^{\epsilon}_{1|0}&=\frac{1}{2}\ket{\phi_{+}^{\epsilon}}\bra{\phi_{+}^{\epsilon}},\\
	\sigma^{\epsilon}_{0|1}&=\frac{1}{2}\ket{1}\bra{1},
	& \sigma^{\epsilon}_{1|1}&=\frac{1}{2}\ket{\phi_{-}^{\epsilon}}\bra{\phi_{-}^{\epsilon}},
\end{align*}
where 
\begin{align*}
\ket{\phi_{+}^{\epsilon}}&=\epsilon\ket{0}+\sqrt{1-\epsilon^2}\ket{1},\\ \ket{\phi_{-}^{\epsilon}}&=\epsilon\ket{1}-\sqrt{1-\epsilon^2}\ket{0}.
\end{align*}
 As long as $\epsilon>0$, this is an extremal assemblage, and thus $E_{\mt{FA}}(\bs{\sigma}^{\epsilon})=S(\mathbb{I}/2)=1$. However, as soon as $\epsilon=0$, the assemblage is decomposable into $\bs{\sigma}^{0}=1/2(\bs{\sigma}^{a}+\bs{\sigma}^{b})$, where
\begin{align*}
	\sigma^{a}_{0|0}&=\ket{0}\bra{0}=\sigma^{a}_{0|1} & \sigma^{a}_{1|0}&=0=\sigma^{a}_{1|1}\\
		\sigma^{b}_{1|0}&=\ket{1}\bra{1}=\sigma^{b}_{1|1} & \sigma^{b}_{0|0}&=0=\sigma^{b}_{0|1}
\end{align*}
and thus $E_{\mt{FA}}(\bs{\sigma}^{0})=0$. This kind of noncontinuity is to be expected: the same example can be used to show the noncontinuity of the steering weight.

In order to prove that extremal points with equal $E_{\mt{FA}}$ do not form the optimal decomposition, we first introduce a parametrisation of  $|\mc{N}|=|\mc{R}|=\mathrm{dim}(\mathcal{H}_B)=2$ assemblages.\\
Using the Bloch vector notation, we can represent the marginal $\rho_{\bs{\sigma}}$ and the matrix $\sigma_{r}:=\sigma_{0|r}-\sigma_{1|r}$ as
\begin{align*}
	\rho_{\bs{\sigma}} &=1/2\left(\begin{array}{cc}
		1+z & x-iy\\
		x+iy & 1-z
	\end{array}\right), \\
 \sigma_{r}&=1/2\left(\begin{array}{cc}
		t_{r}+z_{r} & x_{r}-iy_{r} \\
		x_{r}+iy_{r} & t_{r}-z_{r}
	\end{array}\right).
\end{align*}
In this representation we represent the marginal as a Bloch vector $\mathbf{x}_\sigma$, and each measurement as a four-vector $\tilde{\bs{x}}_r=(t_r,x_r,y_r,z_r)$ which may be separated into a trace component, $t_r$, and a Pauli component, $\mathbf{x}_r=(x_r,y_r,z_r)$. This takes care of the nosignalling condition. Extremal assemblages in this parametrization satisfy
\begin{align*}
	\frac{\n{\mathbf{x} + \mathbf{x}_{r}}+ \n{\mathbf{x} - \mathbf{x}_{r}}}{2} &=1, \\
	\frac{\n{\mathbf{x} + \mathbf{x}_{r}} - \n{\mathbf{x} - \mathbf{x}_{r}}}{2}&=t_r
\end{align*}
where $\mathbf{x}_{0}\neq \mathbf{x}_{1}$. In the degenerate case, $\mathbf{x}=\pm \mathbf{x}_{0}=\pm \mathbf{x}_{1}$ and $\n{\mathbf{x}}=1$.

\begin{figure}
\begin{center}
\includegraphics[scale=0.4]{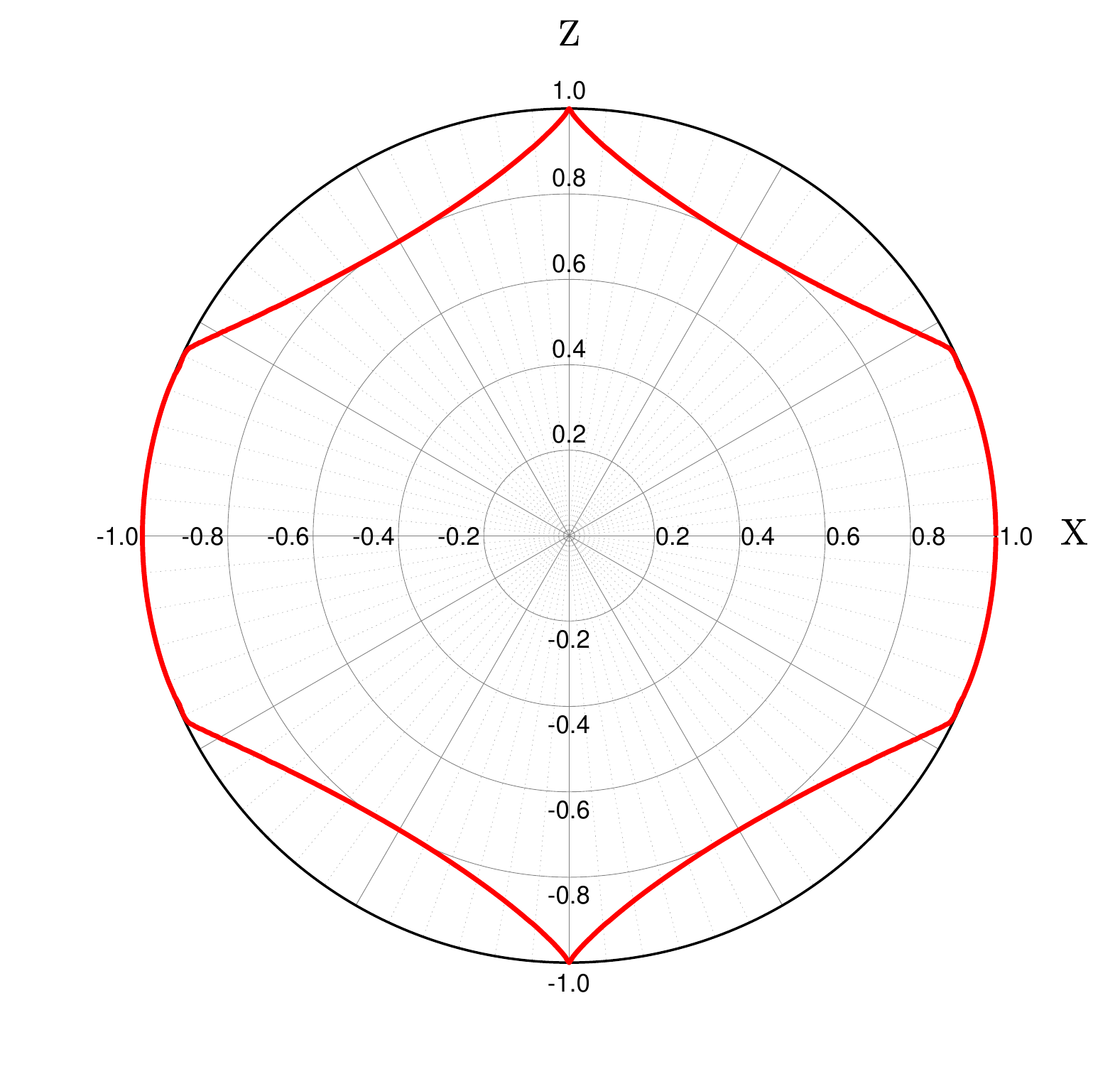}
\caption{Comparison of decompositions for $t_0=0$, $\tilde{\bs{x}}_1=(0,x_1,0,z_1)$ assemblages. Inside the red region equal-value decompositions give a better upper bound, outside the region the steering weight gives the better bound.}\label{PolarComp}
\end{center}
\end{figure}

\begin{figure}
	\begin{center}
		\includegraphics[scale=0.3]{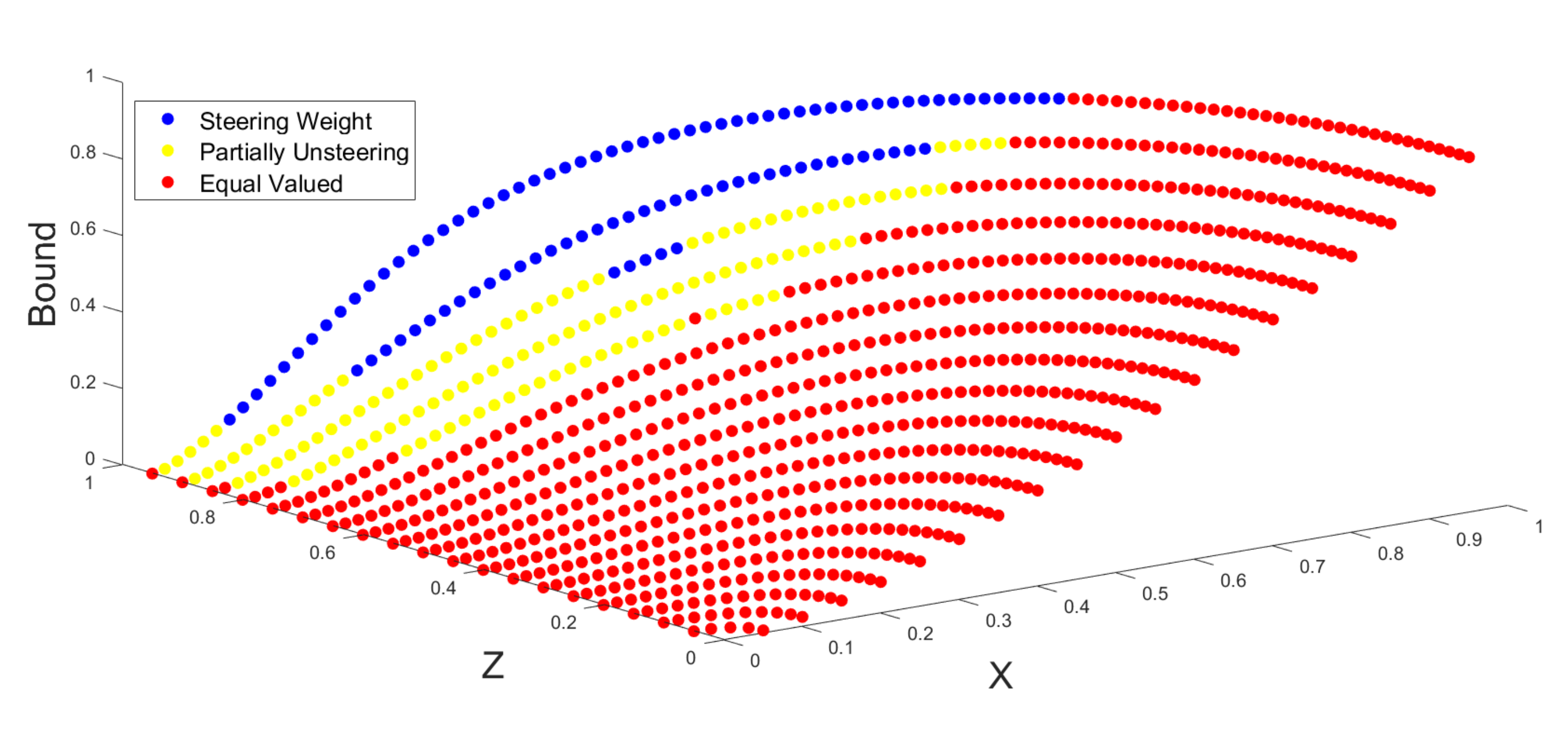}
		\caption{Results of numerical decomposition optimisation for $t_0=0$, $\tilde{\bs{x}}_{1}=(0,x_1,0,z_1)$ assemblages. We find there is a region where neither the equal-value nor steering weight decompositions are optimal.}\label{OptComp}

	\end{center}
\end{figure}

Our entanglement quantity (on extremal states) is then calculated by
\begin{align}
	E_{\mt{FA}}(\bs{\sigma})=E(\mathbf{x}):=&\frac{1+\n{\mathbf{x}}}{2}\log \left(\frac{1+\n{\mathbf{x}}}{2}\right)\nonumber\\
	 +& \frac{1-\n{\mathbf{x}}}{2}\log \left(\frac{1-\n{\mathbf{x}}}{2}\right).
\end{align}
(This is the same formula as the von Neumann entropy for qubits).
We will focus on assemblages with the condition 
\begin{align*}
	\sigma_{0|0} &= \ket{0}\bra{0}/2, & \sigma_{1|0} &= \ket{1}\bra{1}/2.
\end{align*}
Any decomposition of this into extremal points $\bs{\sigma}^{i}$ must have $\sigma^{i}_{0|0} \propto \ket{0}\bra{0}$, $\sigma^{i}_{1|0} \propto \ket{1}\bra{1}$, since these are only valid convex decompositions of rank-1 operators. These extremal assemblages must therefore have $\mathbf{x}^{i}_0=(0,0,1)$ and $\mathbf{x}^{i}=(0,0,t_0)$. Therefore the set of extremal states with constant $E$ have the same value of $|t_0|$. Furthermore the parameters of $\tilde{\bs{x}}_1^{i}$ must satisfy
\begin{align}
	&(x^{i}_1)^2+(y^{i}_1)^2=(1-(z^{i}_1)^2)(1-(t^{i}_0)^2), & t^{i}_1&=z^{i}_1 t^{i}_0. &&\label{ExtLim}
\end{align}
In our restricted subspace, we can therefore describe states by $t_0$ and $\tilde{\bs{x}}_1$.\\
Let us consider an assemblages $\bs{\sigma}$ with $t_0=0$, $\tilde{\bs{x}}_1 = (0,x_1,y_1,z_1)$. For which values of $T\geq 0$ can it be decomposed into extremal points all with $\n{t_0^i}=T$? We need to weight contributions of $t_0^i=T$ and $t_0^i=-T$ equally, and we can always set $t_1=0$ by choosing with equal weight points $t^{i_1}_0=T,\, \tilde{\bs{x}}^{i_1}=(t^{i}_0z_1^{i},x^{i}_1,y^{i}_1,z^{i}_1)$ and $t^{i_2}_0=-T,\, \tilde{\bs{x}}^{i_2}=(-t^{i}_0z^{i},x^{i}_1,y^{i}_1,z^{i}_1)$. Thus from Eq. (\ref{ExtLim}) we can decompose $\bs{\sigma}$ into extremal assemblages with equal $T$ (and thus equal $E$) iff

\begin{equation}
	\frac{(x_1)^2+(y_1)^2}{1-T^2}+(z_1)^2\leq 1. \label{ellipse}
\end{equation} 
As $E$ is monotonically decreasing with the value of $T$, that means the optimal equal-value decomposition is given by the largest $T$ such that the above equation holds. Eq. (\ref{ellipse}) for a fixed $T$ defines a $z$-axis aligned ellipsoid of principal radius 1, and therefore the largest $T$ for which the equation holds is when $(x_1,y_1,z_1)$ lies on the ellipsoid's surface. We name this value of $T$ as $T_{\bs{\sigma}}$, and find that $T_{\bs{\sigma}}=\sqrt{(1-x_1^2-y_1^2-z_1^2)/(1-z_1^2)}$, or $T_{\bs{\sigma}}=1$ if $z_1=1$. The optimal equal-value decomposition is then given by $E(T_{\bs{\sigma}}):=E((0,0,T_{\bs{\sigma}}))$.\\
We can compare the value of this decomposition to that of another explicit decomposition: that provided by the steering weight. This is done in Fig. \ref{PolarComp}. For clarity, we have done this only for \emph{rebit} assemblages, where $y_1=0$. Some further properties of rebit assemblages are given in the supplementary information. From this one can see a clear region, close to the extremal assemblages, in which the steering weight performs better. Thus for these assemblages $E_{\mt{FA}}(\bs{\sigma})\lneq E(T_{\bs{\sigma}})$, which we have already shown to be the lowest valued decomposition for which $E_{\mt{FA}}(\bs{\sigma}^i)$ are all equal.\\
From this argument we are forced to admit the optimal solution cannot consist of equal-valued extremal assemblages; but one could conjecture that by augmenting this strategy with the steering weight decomposition, the optimum is always attained. However, we can again show this is not the case. Fig. \ref{OptComp} shows the result of a numerical optimisation in which an assemblage $\bs{\sigma}$ is optimally decomposed into two further assemblages $\bs{\sigma}^{a},\,\bs{\sigma}^{b}$, which are assigned the value of their equal weight decompositions $E(T_{\bs{\sigma}^{a}})$, $E(T_{\bs{\sigma}^{b}})$. This allows for both the steering weight and equal-weight decompositions. The results of the optimisation show that there is a region with decompositions of a lower value than both $E(T_{\bs{\sigma}})$ and $E_{\mathrm{SW}}(\bs{\sigma})$. These decompositions correspond to the mixing of an unsteering and steering assemblage, where the steering assemblage has higher weight than in the steering-weight achieving decomposition.\\

\section{Conclusions and Discussions}
In this work, we have defined two new quantities for studying the connection between entanglement and steering, by quantifying the amount of entanglement required to create an assemblage. We have shown that both quantities vanish if the considered assemblage is unsteerable.
Both of these values can be upper bounded via the steering weight, which requires the solution of just one semidefinite problem. We also provide a lower bound taken from secret key distillation, which requires only the computation of certain Schatten norms. However, this lower bound could be further optimised with the use of semidefinite programming. The entanglement cost for assemblages, $E_{\mt{CA}}$, has a natural role as a ``proof of prior entanglement" in semi-device-independent systems; the ability to repeatedly generate the assemblage $\bs{\sigma}$ requires that Alice's device must share entanglement with Bob with a rate of at least $E_{\mt{CA}}(\bs{\sigma})$ per assemblage. This may find application in semi-device-independent protocols in testing rounds, for which an analysis of the relation under a finite number of runs would be vital.


Of these two quantities, we believe the entanglement of formation for assemblages to be the easier to work with, as one can apply the results of convex roof extension theory, and exploit its connection to the entanglement of formation for quantum states (shown in Theorem 2). One important problem therefore is to determine the class of assemblages for which $E_{\mt{CA}}=E_{\mt{FA}}$; the analogous equality holds true for a wide class of quantum states, and therefore we expect the same to be true for assemblages. We do not expect it to hold in general, however. In fact, we even expect that the regularisation, $\lim_{k\rightarrow \infty} E_{\mt{FA}}(\bs{\sigma}^{\times k})/k \neq E_{\mt{CA}}(\bs{\sigma})$. This is because the corresponding result for quantum states \cite{HHT2001}, $\lim_{k\rightarrow \infty} E_{\mt{F}}(\rho^{\otimes k})/k = E_{\mt{C}}(\rho)$, relies on the continuity properties of $E_{\mt{F}}$, which we have shown do not transfer to $E_{\mt{FA}}$. Intuitively, an example of nonequality would follow from an optimal distillation protocol in which the \emph{measurement} choices are only asymptotically exact; since our definition $E_{\mt{FA}}$ requires there exact measurements which create exactly the assemblage from a given state.\\
	Another important question is how to calculate $E_{\mt{FA}}$ for general assemblages: a big result here would be to classify for which assemblages the steering weight gives the optimal value, since this is easily calculable. This question is a direct analogy to asking when the ``fully entangled fraction" \cite{BDSW1996} of a state equals its entanglement of formation.\\
One can also imagine performing a similar analysis as is done here, but taking \emph{measurement incompatibility} as the primary resource; asking what are the most compatible measurements that can create a given assemblage. It is already a known result \cite{KSCAA2015,MGHUG2016} that commuting measurements cannot create a steerable assemblage, and vice-versa, that any unsteerable assemblage can be created with commuting measurements. Some work in this direction has already been done \cite{UBGP2015,CS2016}. \\
A natural consequence of this work is that we obtain a \emph{total ordering} on steering assemblages. If $E_{\mt{CA}}(\bs{\sigma}^a)\leq E_{\mt{CA}}(\bs{\sigma}^b)$, then sufficient entanglement to create $\bs{\sigma}^b$ is sufficient to create $\bs{\sigma}^a$. The relation can be used to give a necessity condition about conversion of assemblages via 1-way LOCC (the set of mappings which do not increase steering \cite{GA2015}): if $\bs{\sigma}^a$ can be created by 1-way LOCC from $\bs{\sigma}^b$ then we must have $E_{\mt{CA}}(\bs{\sigma}^a)\leq E_{\mt{CA}}(\bs{\sigma}^b)$. This is because creating first $\bs{\sigma}^b$ from our entanglement, then transforming it to $\bs{\sigma}^a$, gives us a valid sufficient entanglement for $\bs{\sigma}^a$. We note here that such relations have been studied in detail for general resource conversion under Local Operations and Shared Randomness (LOSR) in \cite{SRB2020}.\\
Finally, this work could be extended to continuous variable steering, which is also an active field of research \cite{WJD2007,GHDFFPWS2015,Walketal2015,KBUP2017}.







\begin{acknowledgments}
This work was supported, in part, by the DFG through SFB 1227 (DQ-mat), the RTG 1991, and funded by the Deutsche Forschungsgemeinschaft (DFG, German Research Foundation) under Germany’s Excellence Strategy – EXC-2123 Quantum Frontiers – 390837967. The author would like to thank Tobias Osborne, Reinhard Werner and Ramona Wolf for useful discussions.
\end{acknowledgments}

\bibliographystyle{naturemag}

\appendix
\section{Copies and Distances of Steering Assemblages}
\label{app:assem}
In this section we explain in more detail what it means to have multiple copies of an assemblage, and what distance we are implicitly using when talking about continuity.\\
We consider an assemblage as an ordered tuple of subnormalised density matrices $\boldsymbol{\sigma}:= (\sigma_{0|0},\ldots ,\sigma_{(|\mc{N}|-1)|0},\ldots ,\sigma_{0|(|\mc{R}|-1)},\ldots ,\sigma_{(|\mc{N}|-1)|(|\mc{R}|-1)})$ which satisfy the nosignalling condition, $\sum_{n} \sigma_{n|r} = \rho_{\bs{\sigma}}$. What does it mean to have two copies of an assemblage? Intuitively, an assemblage is a system in which by Alice choosing an input $r$, she can probabilistically steer Bob's state into the collection of substates $\{\sigma_{n|r}\}_{n}$. Thus, with two copies of an assemblage, Alice should be able to choose an $r_1$ and $r_2$, which independently steer Bob's first state into $\{\sigma_{n_1|r_1}\}_{n_1}$ and his second into $\{\sigma_{n_2|r_2}\}_{n_2}$. From this idea, we define two copies of an assemblage as a larger assemblage, written as the ordered tuple
\begin{align}
	\bs{\sigma}^{\times 2}:=(&\sigma_{11|11}\ldots \sigma_{(|\mc{N}|-1)1|(|\mc{R}|-1)1}\ldots \\
	\ldots\, &\sigma_{1(|\mc{N}|-1)|1(|\mc{R}|-1)} \ldots \sigma_{(|\mc{N}|-1)(|\mc{N}|-1)|(|\mc{R}|-1)(|\mc{R}|-1)}),\nonumber
\end{align}
where $\sigma_{n_1n_2|r_1r_2}:=\sigma_{n_1|r_1}\otimes \sigma_{n_2|r_2} \geq 0$. Furthermore, we have that
\begin{equation}
	\sum_{n_1,n_2} \sigma_{n_1n_2|r_1r_2} = \left(\sum_{n_1} \sigma_{n_1|r_1}\right)\otimes \left(\sum_{n_2} \sigma_{n_2|r_2}\right) = \rho_{\bs{\sigma}}\otimes \rho_{\bs{\sigma}}.
\end{equation}
So $\rho_{\bs{\sigma}^{\times 2}}=\rho_{\bs{\sigma}}^{\otimes 2}$.\\

Suppose we have a state and measurements $\rho^{AB}$ and $M^{A}_{n|r}$ such that $\sigma_{n|r}=\mathrm{Tr}_A\left(M^A_{n|r}\otimes \mathbb{I}^B\rho^{AB}\right)$. We then see that
\begin{align*}
	&\mathrm{Tr}_{A_1A_{2}}\left(M^{A_1}_{n_1|r_1}\otimes M^{A_2}_{n_2|r_2}\otimes \mathbb{I}^{B_1B_2}\rho^{A_1B_1}\otimes \rho^{A_2B_2}\right)=\\
	&\mathrm{Tr}_{A_1}\left(M^{A_1}_{n_1|r_1}\otimes \mathbb{I}^{B_1}\rho^{A_1B_1}\right) \otimes \mathrm{Tr}_{A_2}\left(M^{A_{2}}_{n_2|r_2}\otimes \mathbb{I}^{B_2}\rho^{A_2B_2}\right)=\\
	&\sigma_{n_1|r_1}\otimes \sigma_{n_2|r_2}=:\sigma_{n_1n_2|r_1r_2}.
\end{align*}
This fits with our intuition that having two copies of the state and performing separately the correct measurements, we should be able to create two copies of the assemblage. 
The concept described above then generalises to $\times k$. We can describe an element of the assemblage $\bs{\sigma}^{\times k}$ succinctly using $\sigma_{\bf{n}|\bf{r}}$, where $\bf{n},\bf{r}$ are two length-$k$ vectors, where the $i$\textsuperscript{th} entry represents an output or input respectively for the $i$\textsuperscript{th} copy of the assemblage.\\

For the concept of continuity, note that we are describing our assemblages within the more general set of a Cartesian product of finite operators. Therefore, we choose our norm for this space as
\begin{equation}
	\|\bs{\sigma}\|=\sum_{n,r}\|\sigma_{n|r}\|,
\end{equation}
where $\|\sigma_{n|r}\|$ is a norm for finite operators e.g. Hilbert-Schmidt.\\
We can check the three norm properties:
\begin{equation*}
	\|\bs{\sigma}\|=0 \Rightarrow \|\sigma_{n|r}\|=0 \Rightarrow \sigma_{n|r}=0 \Rightarrow \bs{\sigma}=\mathbf{0},
\end{equation*}
\begin{align*}\|\alpha \bs{\sigma}\| = \sum_{n,r}\|\alpha\sigma_{n|r}\|
	=|\alpha|\sum_{n,r}\|\sigma_{n|r}\|
	=|\alpha|\|\bs{\sigma}\|
\end{align*}
and
\begin{align*}
	\|\bs{\sigma}+\bs{\gamma}\|
	&= \sum_{n,r} \| \sigma_{n|r} + \gamma_{n|r}\| \\
	&\leq \sum_{n,r} \left(\| \sigma_{n|r}\| + \|\gamma_{n|r}\|\right)\\ 
	&= \|\bs{\sigma}\|+\|\bs{\gamma}\|.
\end{align*}
Also note that
\begin{align*}
	\|\bs{\sigma}^{\times 2}\|&=\sum_{n_1,n_2,r_1,r_2} \| \sigma_{n_1n_2|r_1r_2}\|\\
	&= \sum_{n_1,n_2,r_1,r_2} \| \sigma_{n_1|r_1} \otimes \sigma_{n_2|r_2} \|\\
	&= \sum_{n_1,n_2,r_1,r_2} \| \sigma_{n_1|r_1}\|\|\sigma_{n_2|r_2} \|\\
	&=\left(\sum_{n_1,r_1} \| \sigma_{n_1|r_1}\|\right) \left(\sum_{n_2,r_2} \| \sigma_{n_2|r_2}\|\right)
	&= \|\bs{\sigma}\|^2.
\end{align*}
where we have assumed a norm is chosen such that $\|A\otimes B\| = \|A\|\|B\|$ such as Hilbert-Schmidt. We use the metric induced by this norm for discussions of continuity.\\

\section{Creation of Steering Assemblages}
The entanglement cost we consider is the following,
\begin{align}
	E_{\mt{CA}}(\bs{\sigma}):= \inf \bigg\{ &E \mid \forall \epsilon > 0, \delta > 0, \exists p,q,\nonumber \\
	&\forall \mathbf{n}\in\mc{N}^{\times q}, \mathbf{x}\in\mc{X}^{\times q},\nonumber\\
	&\exists M_{\bs{n}|\bs{r}}\geq 0, \sum_{\bs{a}} M_{\bs{n}|\bs{r}}=\mathbb{I}^{A}, \Lambda\; \text{and} \nonumber \\
	\sum_{\bs{n},\bs{r}} \big\|\sigma_{\bs{n}|\bs{r}}-\mathrm{Tr}_{A}&\left[\left(M_{\bs{n}|\bs{r}}\otimes \mathbb{I}_{B}\right)\Lambda(\ket{\Phi^+}\bra{\Phi+}^{\otimes p})\right]\big\| \leq \epsilon\bigg\}.
\end{align}
Similar to the entanglement cost for quantum states, we allow for the possibility of a cheaper cost when preparing many copies of the assemblage, and also allow for protocols which are only asymptotically perfect in creating the assemblage. Note that these imperfections may be in the state, or in the measurements. Although the state before measurement can be quite general (reflected by e.g. the $\mathbb{I}^{A}$, denoting an arbitrary sized system for Alice before measurement) it should result in an assemblage with the same input/output set as $\sigma_{\bs{n}|\bs{r}}$ and the same Hilbert space dimension, in order for the distance comparison to be valid.

A reasonable question to ask is whether this definition should allow for some LOCC operations \emph{after} the measurement; which itself is an LOCC operation. We believe that is not necessary, and that all operations possible after the measurement can be absorbed into $\Lambda$, or the definition of the measurement operators. We will argue this below.\\

Another point of consideration is, if LOCC operations are allowed after the measurement operator, are all LOCC operations allowed? Here we argue that the answer to this second question is no. Let us suppose that once an assemblage is created, it is distributed to two parties (Alexa and Boris) who use it for a semi-device independent protocol. Before Alexa has chosen an input $r$, they can ensure that no signalling can enter or leave their laboratories. Therefore, we do not allow distillation procedures in which Alice's system signals information about $n$ or $r$, after the input is chosen and output determined (done by $M_{n|r}$). The device may only be prepared to locally react to input choices and output results, explained by the \emph{wirings} given below.  We now show that all other allowed operations can be absorbed into $\Lambda$, or the definition of the measurement operators.
\begin{itemize}
	\item \emph{Sharing of randomness for joint operations and deciding which operations to be done.}\\
	All of this can be shared and communicated during the LOCC protocol $\Lambda$, before the measurement takes place.\\
	\item \emph{Tracing out of extra systems, and operations independent of $n,r$ and $\rho_{\mathbf{\sigma}}$}.\\
	Since these either commute with the measurement operators, or do not require the assemblage to exist when performing them, these can also be moved into the $\Lambda$ protocol before the measurements.\\
	\item \emph{Bob performs an operation on his state, and communicates the result to Alice.}\\
	Since Alice's measurements are performed locally, they commute with Bob's local operations, which can take place beforehand. So this is also absorbed into $\Lambda$.\\
	\item \emph{Alice rewires her possible inputs.}\\
	Given Alice has access to input choices in $\mc{R}$, a rewiring of inputs corresponds to a conditional probability distribution $p(r|\Omega,r')$, with $r'$ in a (potentially) different input choice set $\mc{R}'$. Alice will then choose an input $\mc{R'}$, which will, along with the outcome of some random variables $\Omega$, determine which choice is inputted into the original assemblage. As discussed, any generation of $\Omega$ can be absorbed into the $\Lambda$ beforehand; meanwhile, a rewiring can be absorbed into the measurements by rewriting
	$M_{n|r'}=\sum_{r} p(r|\Omega, r')M_{n|r}$. This is again positive and moreover we have that
	$\sum_{n} M_{n|r'} = \sum_{n,r}p(r|\Omega, r')M_{n|r} = \sum_{r}p(r|\Omega, r')\sum_{n}M_{n|r} =\sum_{r}p(r|\Omega, r')\mathbb{I}^{A_{p}}$. Since we are considering LOCCs and not more general stochastic LOCCs (SLOCCs), this wiring must be deterministic (normalised implying $\sum_{r}p(r|\Omega, r'=1)$) and so we see that $r'$ remains a valid measurement.\\
	\item \emph{Alice rewires her possible outputs.}\\
	Similarly, Alice may decide to implement a conditional probability distribution $p(n'|\Omega,r',r,n)$, which receives output $n$ from the original assemblage and maps it to $n'$, an output in a (possibly different) output set $\mc{N}'$. We may in a similar manner define
	$M_{n'|r}= \sum_{n}p(n'|\Omega,r',r,n)M_{n|r}$. This again gives us positive operators, where
	$\sum_{n'} M_{n'|r} = \sum_{n}\sum_{n'} p(n'|\Omega,r',r,n)M_{n|r} = \sum_{n} M_{n|r} = \mathbb{I}^{A_p}$, where we have again used the deterministic nature of the wirings. Thus, we have incorporated our wiring into the new measurement $\{M_{n|r}\}$.\\
\end{itemize}

\section{Rebit Assemblages}\label{app:rebits}
In this appendix we talk shortly about, when calculating the $E_{\mt{CA}}$ for an assemblage with $\sigma_{0|0}=\ket{0}\bra{0}/2$, $\sigma_{1|0}=\ket{1}\bra{1}/2$,  how one can ensure that the matrix $\sigma_{1}=\sigma_{0|1}-\sigma_{1|1}$ can always be taken to be real. To see this, we note that the unitary $U_{\theta}=\cos{\theta}\sigma_x + \sin{\theta}\sigma_y$ (here $\sigma_x,\sigma_y$ are the Pauli matrices) transforms the matrix $\sigma_{1}$ as
\begin{equation}\label{rebtrans}
	U_{\theta}\sigma_{1}U^{\dagger}_{\theta} = 1/2\left(\begin{array}{cc}
		t_1 + z_1 & e^{-2i\theta}(x_1-iy_1) \\
		e^{2i\theta}(x_1+iy_1) & t_1-z_1
	\end{array}\right).
\end{equation}
Thus, by choosing $\theta=\mathrm{Arg}(x_1-iy_1)/2$, this transformation takes it to a real matrix with $x_1'=\sqrt{x_1^2+y_1^2}$ and $y_1'=0$.\\
To see that this transformation can be realised, consider what happens if Bob performs this unitary on his local system. Since we constrained $\sigma_{0|0} = \ket{0}\bra{0}/2$ and  $\sigma_{1|0} = \ket{1}\bra{1}/2$, these are left invariant by the unitary, as is the marginal $\rho_{\bs{\sigma}}=\sigma_{0|0}+\sigma_{1|0}$, since it is diagonal in the Pauli-$Z$ basis. However, $\sigma_{0|1}\rightarrow \sigma'_{0|1}=U_{\theta}\sigma_{0|1}U^{\dagger}_{\theta}$ and $\sigma_{1|0}\rightarrow \sigma'_{1|1}=U_{\theta}\sigma_{1|1}U^{\dagger}_{\theta}$ . Therefore,
\begin{equation}
	\sigma'_{1}=\sigma'_{0|1}-\sigma'_{1|1}  =U_{\theta}(\sigma_{0|1}-\sigma_{1|1})U^{\dagger}_{\theta} = U_{\theta}\sigma_{1}U^{\dagger}_{\theta},
\end{equation}
and so we see that the transformation in Eq. (\ref{rebtrans}) is a physically realisable one. We also note that such a unitary commutes with Alice's measurement (since they take place on different subsytems), and any state used to create the first assemblage is locally unitarily equivalent to any to create the second assemblage. Since entanglement measures are invariant under local unitaries, we can conclude that such local unitaries do not affect the value of $E_{\mt{CA}}/E_{\mt{FA}}$.\\
We also make the comment here that the optimal decomposition of a rebit assemblage must be into extremal rebit assemblages. Let us assume the optimal decomposition contains the extremal point $\bs{\sigma}^i$ with weight $p^{i}$, whose parameters are $t^{i}_0$ and $\tilde{\bs{x}}^{i}_1=(t^{i}_0z^{i}_1,x_1^{i},y_{1}^{i},z_{1}^{i}),\;y_{1}^{i}\neq 0$. As our assemblage is invariant under the transformation $y_1\rightarrow -y_1$, so too must our optimal decomposition, and therefore it must contain the assemblage $\bs{\sigma}^{j}$, also with weight $p_i$, whose parameters are $t^{j}_0=k^{i}_0$ and $\tilde{\bs{x}}^{j}_1=(t^{i}_0z^{i}_1,x_1^{i},-y_{1}^{i},z_{1}^{i})$.
Together these two points contribute, with weight $2p_i$, the assemblage with $t_0=t^{i}_0$, 
$\tilde{\bs{x}}_1=(t^{i}_0z^{i}_1,x_1^{i},0,z_{1}^{i})$.
However, there exists an extremal point (which we call $\bs{\sigma}^{i'}$) with $|t^{i'}_0|>|t^{i}_0|$ and $\tilde{\bs{x}}^{i'} = (t^{i'}_0z_{1}^{i},x_1^{i},0,z_{1}^{i})$ (this follows from Eq. (25) in the main body). Furthermore, the point  $\bs{\sigma}^{j'}$ with $t^{j'}_0=-t^{i'}_0$ and $\tilde{\bs{x}}^{j'} = (-t^{i'}_0z_{1}^{i},x_1^{i},0,z_{1}^{i})$ is also extremal. By mixing together these two points in the ratio $(1+t_0^{i})/(2t^{i'})\bs{\sigma}^{i'} + (1-t_0^{i})/(2t^{i'})\bs{\sigma}^{j'}$ we find we create the same assemblage as $1/2\bs{\sigma}^{i}+1/2\bs{\sigma}^{j}$, and can therefore replace $\bs{\sigma}^{i},\bs{\sigma}^{j}$ in the decomposition with $\bs{\sigma}^{i'},\bs{\sigma}^{j'}$.
\pagebreak

As we have $E(\mathbf{x}^{i'})<E(\mathbf{x}^{i})$ and $E(\mathbf{x}^{j'})<E(\mathbf{x}^{j})$ (due to the monotonicity of $E$) our original decomposition was not optimal -- this is a contradiction. Thus, we can conclude that $y_1^{i}=0$ for all extremal $\bs{\sigma}^{i}$ in the optimal decomposition.

\section{Numerical Algorithm}
In this section we explain briefly how the numerical optimisation algorithm works, and how it incorporates the equal value decomposition and steering weight into the optimisation.\\
The algorithm uses MATLAB's \emph{fmincon} function to split an assemblage $\bs{\sigma}=p\bs{\sigma}^{a} + (1-p)\bs{\sigma}^{b}$, in order to minimise the quantity $p E(T_{\bs{\sigma}^a})+(1-p)E(T_{\bs{\sigma}^b})$. By setting $p=1$ and $\bs{\sigma}^{a}=\bs{\sigma}$, we see the equal value decomposition is considered by this optimisation.\\
In order to prove that this optimisation can also achieve the steering weight decomposition, we need to show that for an assemblage in our restricted subspace with $t_0=0=t_1$, there exists a decomposition 
$\bs{\sigma} = (1-SW(\bs{\sigma}))\bs{\sigma}^{\mathrm{LHS}}+SW(\bs{\sigma})\bs{\gamma}$,
where $t^{\mathrm{LHS}}_0=t^\mathrm{LHS}_1=0$.
To do this, we first rewrite the steering weight in an equivalent form
\begin{equation}
	1-SW(\bs{\sigma})=\max\{\mathrm{Tr}(\bs{\sigma}^{\mathrm{LHS}})\mid \bs{\sigma}=\bs{\sigma}^{\mathrm{LHS}}+\bs{\gamma}  \},
\end{equation}
where now $\bs{\sigma}^{\mathrm{LHS}},\bs{\gamma}$ are subnormalised assemblages, i.e. $\mathrm{Tr}(\rho_{\bs{\sigma}^{\mathrm{LHS}}}),\mathrm{Tr}(\rho_{\bs{\gamma}})\leq 1$.\\
Next, we note that due to our restricted subspace, we are limited to assemblages with $\sigma_{0|0}\propto \ket{0}\bra{0}$, $\sigma_{1|0}\propto \ket{1}\bra{1}$. This means our unsteerable assemblage must be a convex combination of the following four assemblages:
\begin{align*}
	\sigma^{1}_{0|0}&=\ket{0}\bra{0}=\sigma^{1}_{0|1} &  \sigma^{1}_{1|0}&=0=\sigma^{1}_{1|1},\\
	\sigma^{2}_{0|0}&=\ket{0}\bra{0}=\sigma^{2}_{1|1} &  \sigma^{2}_{1|0}&=0=\sigma^{2}_{0|1},\\
	\sigma^{3}_{1|0}&=\ket{1}\bra{1}=\sigma^{3}_{0|1} &  \sigma^{3}_{1|0}&=0=\sigma^{3}_{1|1},\\
	\sigma^{4}_{1|0}&=\ket{1}\bra{1}=\sigma^{4}_{1|1} &  \sigma^{4}_{1|0}&=0=\sigma^{4}_{0|1}.\\
\end{align*}
Therefore any $\bs{\sigma}^{\mathrm{LHS}}$ must be of the form $\bs{\sigma}^{\mathrm{LHS}}=\sum_{i} \alpha^i \bs{\sigma}^i$, $\alpha^i\geq 0$, $\sum_{i} \alpha^{i}\leq 1$. In particular, 
\begin{align*}
	t^{\mathrm{LHS}}_0&=\alpha^1+\alpha^2-\alpha^3-\alpha^4, & t^{\mathrm{LHS}}_1&=\alpha^1+\alpha^3-\alpha^2-\alpha^4.
\end{align*}
The steering weight problem is now equivalent to $\max \sum_{i} \alpha^i$ such that
\begin{align*}
	&(1/2-\alpha^1-\alpha^2)\ket{0}\bra{0}\geq 0,\\
	&(1/2-\alpha^3-\alpha^4)\ket{1}\bra{1}\geq 0,\\
	&\sigma_{0|1}-\alpha^1\ket{0}\bra{0}-\alpha^3\ket{1}\bra{1}\geq 0,\\
	&\sigma_{1|1}-\alpha^2\ket{0}\bra{0}-\alpha^4\ket{1}\bra{1}\geq 0.\\
\end{align*}


Using the fact that a 2x2 matrix is positive semidefinite iff its trace and determinant are nonnegative, and that our original assemblage can be explicitly expressed in our parameterisation as
\begin{align*}
	\sigma_{0|1}&=1/4\left(\begin{array}{cc}
		1+z_1 & x_1-iy_1 \\
		x_1+iy_1 & 1-z_1 
	\end{array}\right)\\
	\sigma_{1|1}&=1/4\left(\begin{array}{cc}
		1-z_1 &-x_1+iy_1 \\
		-x_1-iy_1 & 1+z_1
	\end{array}\right)
\end{align*}

We can rewrite the optimization as $\max  \sum_i \alpha^i$
such that
\begin{align*}
	&(1/2-\alpha^1-\alpha^2)\geq 0,\\
	&(1/2-\alpha^3-\alpha^4)\geq 0,\\
	&(1/2-\alpha^1-\alpha^3)\geq 0,\\
	&(1/2-\alpha^2-\alpha^4)\geq 0,\\
	&(1+z_1-4\alpha^1)(1-z_1-4\alpha^3)-x_1^2-y_1^2\geq 0,\\
	&(1+z_1-4\alpha^4)(1-z_1-4\alpha^2)-x_1^2-y_1^2\geq 0.\\
\end{align*}
We can see that this problem is invariant under the transformations $\alpha^1\leftrightarrow \alpha^4$, $\alpha^2\leftrightarrow \alpha^3$. Thus, if there exists an optimal solution $\alpha^1=\beta^1,\,\alpha^2=\beta^2,\,\alpha^3=\beta^3,\,\alpha^4=\beta^4$, then $\alpha^1=\beta^4,\,\alpha^2=\beta^3,\,\alpha^3=\beta^2,\,\alpha^4=\beta^1$ is also an optimal solution. Furthermore, an equal mixture of the two $\alpha^1=\alpha^4=(\beta^1+\beta^4)/2,\,\alpha^2=\alpha^3=(\beta^2+\beta^3)/2$ is also an optimal solution. Moreover, it is an optimal solution with $t^{\mathrm{LHS}}_0=t^{\mathrm{LHS}}_1=0$. This means that it is eligible to be considered in our MATLAB optimisation, and consequently our optimisation will achieve a value less than or equal to $E_{\mathrm{SW}}(\bs{\sigma})$.

\end{document}